\documentclass{article}
\usepackage{graphicx}
\usepackage{amssymb}
\usepackage{amsbsy}
\usepackage{amsfonts}
\usepackage{amsgen}
\begin{document}
\title{The Origin of Helicity in the Solar Active Regions}
\author{Arnab Rai Choudhuri$^{1}$, Piyali Chatterjee$^{2}$, 
Dibyendu Nandy$^{3}$}
\maketitle

{\centering{$^{1, 2}$Department of Physics, Indian Institute of Science, Bangalore-560012, India. \\email: arnab@physics.iisc.ernet.in, piyali@physics.iisc
.ernet.in\\}
\centering{$^{3}$Department of Physics, Montana State University, Bozeman, MT 59717, USA.
\\ \hspace{5cm} email: nandi@mithra.physics.montana.edu}}
\maketitle
\begin{abstract}
We present calculations of helicity based on our solar dynamo model
and show that the results are consistent with observational data.
\end{abstract}

\section{Introduction}

It is known that solar active regions usually have some helicity
associated with them.  Several authors established this from
magnetogram studies of active regions (Seehafer 1990; Pevtsov,
Canfield \& Metcalf 1995; Abramenko, Wang \& Yurchishin 1997;
Bao \& Zhang 1998; Pevtsov, Canfield \& Latushko 2001).  There
are different ways of describing helicity mathematically,  one
of which is
$$\alpha = \frac{(\nabla \times {\bf B})_z}{B_z}, \eqno(1)$$
where $z$ corresponds to the vertical direction.
\begin{figure}
\centerline{\includegraphics[height=6cm,width=7cm]{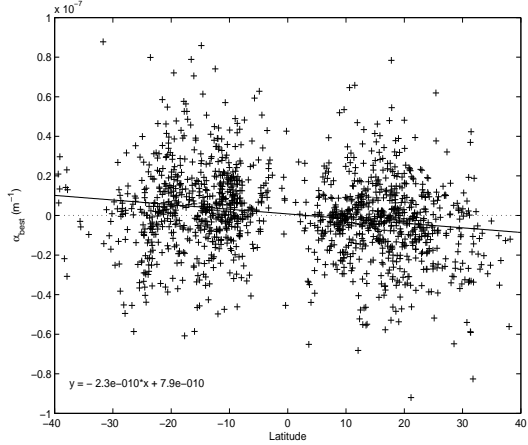}}
\caption{Helicity $\alpha$ for active regions at different latitudes.
Based on data for the duration 1985--2002.}
\end{figure}
Since $\nabla \times {\bf B}$ will be non-zero only if the magnetic
field has a helical structure, $\alpha$ is a good measure of helicity.
Fig.~1 presents observational data showing values of $\alpha$ for
active regions which emerged at different latitudes. It is clear
that there is a preference for negative helicity in the northern
hemisphere and positive helicity in the southern hemisphere.  The
preference is at about 70\% level.

\section{Basic idea}

Our theoretical calculations are based on the solar dynamo model
developed by Nandy \& Choudhuri (2002) and Chatterjee, Nandy \&
Choudhuri (2004).  Here we touch upon the bare essentials of the
model. The toroidal field is produced within the high-latitude
tachocline, where helioseismology has discovered strong shear.
Then the toroidal field is advected by the meridional circulation
to lower latitudes where it enters the convection zone and becomes
buoyant.  Magnetic buoyancy makes the toroidal flux tubes rise to
the solar surface where the poloidal field is generated by the
Babcock--Leighton mechanism, i.e.\ from the decay of tilted bipolar
regions.  

Dynamo models deal with mean fields, whereas we want to find 
helicities of active regions which form from flux tubes.  To make
a connection between these two, we have to look at the relation
between dynamo theory and flux tubes.  This has been explored by
Choudhuri (2003), who presents the qualitative idea of how the helicity
is generated. Fig.~2 shows a section of the convection zone {\em in the
northern hemisphere}, with the pole towards left and the equator towards
right.  Suppose the flux tubes at the bottom have magnetic field into
the paper.  Keeping in mind that the leading sunspot is found near
the equator, it is easy to see that the rising flux tubes will produce
a clockwise poloidal field by the Babcock--Leighton mechanism.  
Now consider a flux tube rising in this region of existing poloidal
field.  Since the magnetic field is nearly frozen, the flux tube
will drag the poloidal field lines which get wrapped around the tube
in anti-clockwise sense. This implies a current out of the paper, 
whereas the magnetic field inside the tube is into the paper---giving
a negative helicity.
\begin{figure}
\centerline{\includegraphics[height=7cm,width=9cm]{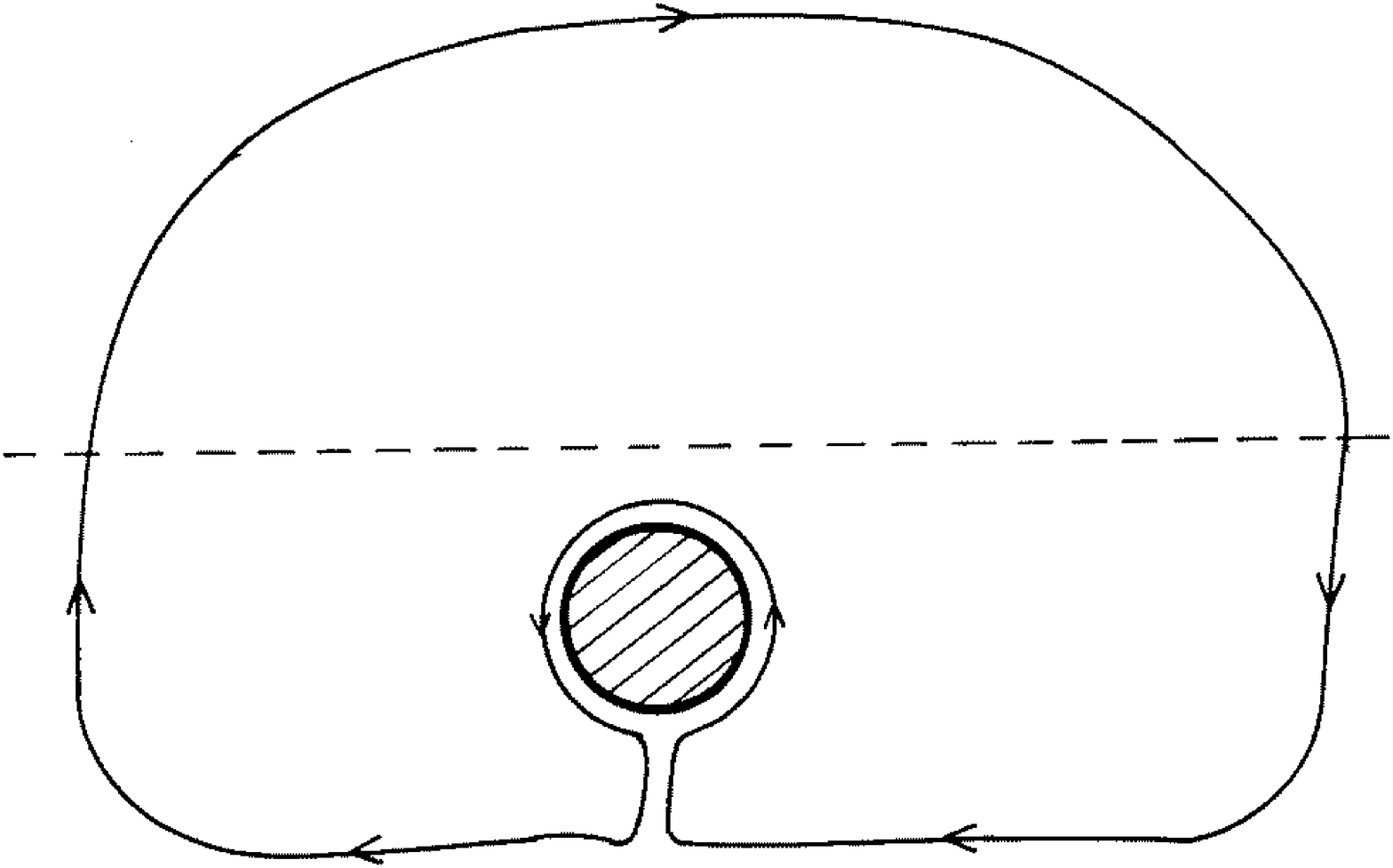}}
\caption{Magnetic field lines around a rising flux tube, with the dashed line indicating the 
solar surface. The flux tube shown by the shaded circle has toroidal field going into the paper,
and is rising through a region of clockwise poloidal field lines.}
\end{figure}

To estimate the magnitude of helicity, keep in mind that the flux
of poloidal field $B_P$ through the whole convection zone gets dragged
by the rising flux tube.  If $d$ is the depth of the convection zone,
the flux dragged by the tube is 
$$F \approx B_P  d. \eqno(2)$$
If this flux $F$ gets wrapped around the tube of radius $a$, then the
magnetic field around the tube is of order $F/a$.  The current 
$|\nabla \times {\bf B}|$ associated with this field is of order
$F/a^2$ and is along the axis of the tube.  If $B_T$ is the magnetic
field inside the flux tube, then it follows from (1):
$$\alpha \approx \frac{F/a^2}{B_T} \approx \frac{B_P d}{B_T a^2} \eqno(3)$$
on substituting from (2) for $F$. On using $B_P \approx 1$ G, the field
inside sunspots $B_T \approx 3000$ G and the radius of the sunspot $a
\approx 2000$ km, we get $\alpha \approx 0.2 \times 10^{-7}$ m$^{-1}$.
Thus, from very simple arguments, we get the correct order of magnitude.

\section{Results from simulation}

For simplicity, let us assume that all flux tubes have the same radius 
$a$.  It then follows from (3) that the helicity of the flux tube is
essentially given by the flux $F$ through the convection zone. Hence, whenever an eruption takes
place in our dynamo simulation, we calculate the poloidal flux through
the convection zone by integrating $B_{\theta}$ in the radial direction
at that latitude.  This gives the helicity associated with the eruption.

Fig.~3 shows the simulated butterfly diagram, indicating where the 
helicity is positive and where it is negative. During the solar maximum,
the helicity is negative in the northern hemisphere and positive in the
southern, as we expect.  However, at the beginning of a cycle, there is
a short duration when the helicity is `wrong'.  Bao, Ai \& Zhang (2000)
reported the evidence of such `wrong' helicity at the beginning of 
Cycle~23.  Basically, when a flux tube erupts
in a region where the poloidal
field has been created by {\em similar} flux tubes which erupted
earlier, we get the correct helicity, as can be seen from Fig.~2.
At the beginning of a cycle, however, flux tubes emerge in regions
where the poloidal field was produced by flux tubes of the earlier
cycle, thereby giving rise to opposite helicity.
\begin{figure}
\centerline{\includegraphics[height=7cm,width=9cm]{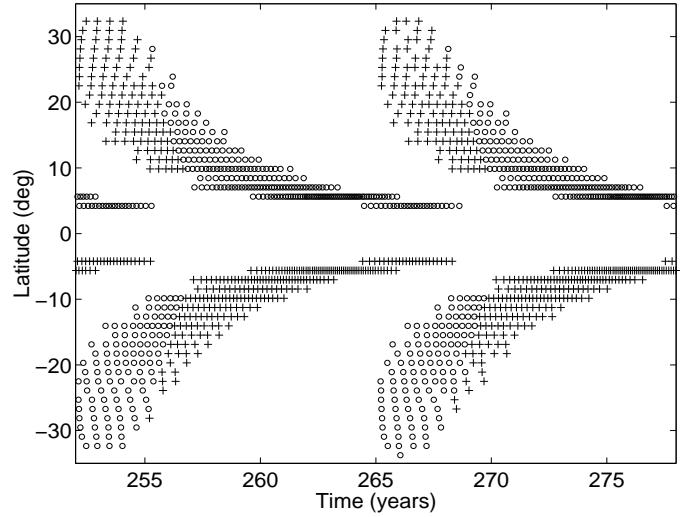}}
\caption{Theoretical butterfly diagram of eruptions from our dynamo 
simulations. Eruptions with positive and negative helicities are
denoted by '+' and 'o' respectively.}
\end{figure}
Finally Fig.~4 is a plot of helicity associated with eruptions at different
latitudes.  In our dynamo model, so far we have not introduced
fluctuations, which is needed to model irregularities of the solar
cycle.  We are now in the process of introducing fluctuations in our
dynamo model, which should increase the scatter in the helicity plot.
Even now, Fig.~4 is not a bad theoretical match for the observational
Fig.~1.
\begin{figure}
\centerline{\includegraphics[height=6cm,width=7cm]{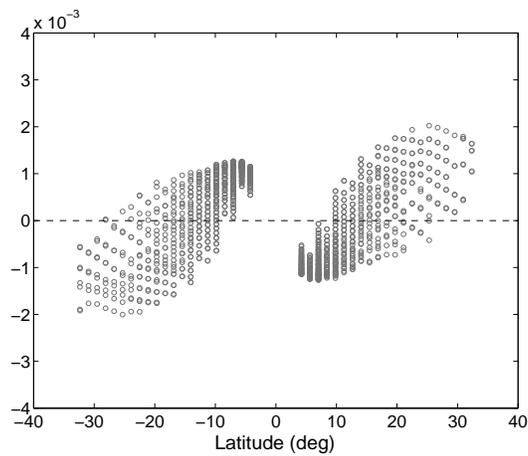}}
\caption{Helicity (in arbitrary units) for eruptions at different latitudes. 
The hemispheric preference is found to hold for about 67\% of the eruptions. 
Compare this theoretical plot with Fig.~1.}
\end{figure}

\section{Conclusion}

Two clear theoretical predictions follow from our model. 
\begin{enumerate}
\item Since
the helicity goes as $a^{-2}$ as seen from (3), the smaller sunspots
should statistically have stronger helicity. 
\item At the beginning
of a cycle, helicity should be opposite of what is usually observed.
\end{enumerate}

There is an alternative model for the generation of helicity: the
$\Sigma$-effect proposed by Longcope, Fisher \& Pevtsov (1998).  This
model also makes the prediction (a), but not (b).  According to
this model, the helicity should not vary with the solar cycle.
As already noted, Bao, Ai \& Zhang (2000) noticed indications of
opposite helicity at the beginning of Cycle~23, which lends support
to our model.  More careful analysis of observational data is needed 
to establish any possible cycle-dependence of helicity.  We hope
that this will be done in near future.   
\section*{References}
Abramenko, V. I., Wang, T., \& Yurchishin, V. B., 1997, \textit{Solar Phys.,} 
\textbf{174}, 291.\\
Bao, S. \& Zhang, H., 1998, \textit{ApJ,} \textbf{496}, L43.\\
Bao, S. Ai, G. X., \& Zhang, H., 2000, \textit{J. Astrophys. Astron.,} 
\textbf{21}, 303.\\
Chatterjee, P., Nandy, D. \& Choudhuri, A. R., 2004,  \textit{A\&A}, 
in press (astro-ph/0405027). \\
Choudhuri, A.R., 2003, \textit {Solar Phys.,} \textbf{215}, 31.\\
Longcope, D. W., Fisher, G. H. \& Pevtsov, A.A., 1998,
\textit{ApJ}, \textbf{507}, 417.\\
Nandy, D. \& Choudhuri, A. R., 2002, \textit{Science}, \textbf{296}, 1671.\\
Pevtsov, A. A., Canfield, R. C., \& Metcalf, T. R., 1995,
\textit{ApJ,} \textbf{440}, L109.\\
{Pevtsov, A. A., Canfield, R. C., \& Latushko, S. M.,} 2001
\textit{ApJ,} \textbf{549}, L261.\\
Seehafer, N., 1990, 
\textit {Solar Phys.,} \textbf{125}, 219.\\

\end{document}